\documentclass[12pt]{article}
\usepackage{graphicx}
\newlength{\vshift}
\input epsf.tex
\newlength{\hshift}

\usepackage{amssymb}
\usepackage{amsmath}
\usepackage{amsthm}
\usepackage{amsopn}
\def\uno{\mbox{1 \kern-.59em {\rm l}}}

\def\beq{\begin{equation}}
\def\eeq{\end{equation}}
\def\bea{\begin{eqnarray}}
\def\eea{\end{eqnarray}}

\begin{document}

 \vspace*{3cm}

\begin{center}

{\bf{\Large    Ruppeiner Geometry of Anyon Gas}}

\vskip 4em

{ {\bf Behrouz ~Mirza} \footnote{e-mail: b.mirza@cc.iut.ac.ir}\:
and \: {\bf Hosein ~Mohammadzadeh}\footnote{e-mail:
h.mohammadzadeh@ph.iut.ac.ir}}

\vskip 1em

Department of Physics, Isfahan University of Technology, Isfahan,
84156-83111, Iran
 \end{center}

 \vspace*{1.9cm}

\begin{abstract}
We derive the thermodynamic curvature of a two dimensional ideal
anyon  gas of particles obeying fractional statistics. The
statistical interactions of anyon gas can be attractive or
repulsive. For attractive statistical interactions, thermodynamic
curvature is positive and for repulsive statistical interactions,
it is negative, which indicates a more stable anyon gas. There is
a special case between the two where the thermodynamic curvature
is zero. Small deviations from the classical limit will also be
explored.
\end{abstract}

PACS number(s): 05.20.-y, 67.10.Fj
%%%%%%%%%%%%%%%%%%%%%%%%%%%%%%%%%%%%%%%%%%%%%%%%%%%%%%%%%%%%%%%%%%%%%%%%%%%%%%%%%%%%%%%%%%%%%%%%%%%%%%%%%%%%%%%%%%%%%%%%%%%
\newpage
\section{Introduction}
In 1979, Ruppeiner introduced a Riemannian metric structure
representing thermodynamic fluctuation theory, which was  related
to the second derivatives of the entropy \cite{Ruppeiner1,
Ruppeiner01}. One of the most significant aspects of his theory
was the introduction of the Riemannian thermodynamic curvature as
a qualitatively new tool for the study of fluctuation phenomena.
On a purely phenomenological level, it was initiated by Weinhold
who introduced a sort of Riemannian metric into the space of
thermodynamic parameters \cite{weinhold1}. However, it turned out
that the two metrics introduced by Weinhold and Ruppeiner were
conformally equivalent \cite{Mrugala1, salamon}. It is natural to
calculate the thermodynamic curvature for models whose
thermodynamics is  exactly known. This has been done by several
authors \cite{Nulton,Ruppeiner2,
Ruppeiner4,Mrugala2,Mrugala3,Brody,kaviani,Johnston1,Johnston2,Ruppeiner3,Rup,Mirza}.
Janyszek and Mruga{\l}a worked out the thermodynamic curvature for
ideal Fermi and Bose gases and reported that the sign of the
thermodynamic curvature is always different for ideal Fermi and
Bose gases. It was argued that the scalar curvature could be used
to show that fermion gases were more stable than boson gases
\cite{Mrugala2}.

For a two dimensional system, the statistical distribution may
interpolate between bosons and fermions when there is no mutual
statistics and that respects a fractional exclusion principle.
Anyons constitute such a physical system and in the present
paper, we investigate the Ruppeiner geometry of an ideal anyon
gas and its stability.

The outline of this paper is  as follows. In Section 2, the
thermodynamic properties of anyons is summarized and the internal
energy for the anyon gas is derived . In Section 3, the Ruppeiner
metric of the parameter space of this system is obtained and,
finally, the thermodynamic curvature of the anyon gas in the
classical limit is evaluated. As we will see, the sign of the
thermodynamic curvature is not constant and a stability condition
can be introduced. In Section 4, the Ruppeiner curvature for small
deviations from the classical limit is considered.

%*************************************************%
\section{Thermodynamic properties of ideal gas of fractional statistical particles }
%*************************************************%
The concept of "anyons" or particles with fractional statistics in
two-dimensional systems \cite{leinaas, wilczek} has found
applications in the theory of fractional quantum Hall effect
\cite{halperin}. Therefore, such particles and their thermodynamic
properties have been the subject of research by a number of
authors \cite{Haldane,Wu,Aoyama,Wung1,Wung2}. The statistical
weight of $N$ identical particles occupying a group of $G$ states
for bosons or fermions is, respectively, given by
    \bea
    W_{b}=\frac{(G+N-1)!}{N!(G-1)!}\:\:\:\:\:\:or\:\:\:\:\:\:
    W_{f}=\frac{G!}{N!(G-N)!}
    \eea
A simple interpolating function which implies fractional exclusion
is
    \bea
    W=\frac{[G+(N-1)(1-\alpha)]!}{N![G-\alpha N-(1-\alpha)]!}
    \eea
with $\alpha=0$ corresponding to bosons, $\alpha=1$ to fermions, and
$0<\alpha<1$ to intermediate statistics. Haldane \cite{Haldane}
defined the statistical interactions $\alpha_{ij}$ through the
linear relation
    \bea
    \Delta d_{i}=-\sum_{i}\alpha_{ij}\Delta N_{j},
    \eea
where $\{\Delta N_{j}\}$ is a set of  changes allowed to occur in
the particle number. $W(\{N_{i}\})$, the number of system
configuration corresponding, to the set of occupation number
$\{N_{i}\}$, is given by
    \bea
    W(\{N_{i}\})=\prod_{i}\frac{[G_{i}+N_{i}-1-\sum_{j}\alpha_{ij}(N_{j}-\delta_{ij})]!}
    {N_{i}![G_{i}-1-\sum_{j}\alpha_{ij}(N_{ij}-\delta_{ij}))]!}
    \eea
The parameter $\alpha_{ij}$ is rational. We call $\alpha_{ij}$ for
$i\neq j$ mutual statistics. The above equation applies to the usual
Bose or Fermi ideal gas with $i$ labeling single particle energy
levels. So with an extension of the meaning of species, this
definition allows different species indices to refer to particles of
the same kind but with different quantum numbers.

Under the constraint of fixed particle number and energy,
    \bea
    N=\sum_{i}N_{i},\nonumber\\E=\sum_{i}\epsilon_{i}N_{i},
    \eea
the grand partition function $Z$ is determined by Haldane and Wu
\cite{Haldane, Wu}  who state the counting rule as follows
    \bea
    Z=\sum_{\{N_{i}\}}W(\{N_{i}\})\exp\{\sum_{i}N_{i}(\mu_{i}-\epsilon_{i})/kT\},
    \eea
where, $\mu $ and $T$ are the Lagrange multipliers incorporating
the constraints of fixed particle number and energy, respectively.

The stationary condition of the grand partition function $Z$ with
respect to $N_{i}$ gives the statistical distribution
$n_{i}\equiv\frac{N_{i}}{G_{i}}$ of an ideal gas of fractional
statistically identical particles with the same chemical potential
$\mu=\mu_{i}$, $\alpha_{ij}=\alpha\delta_{ij}$ and temperature $T$
as derived by Wu,
    \bea
    \label{n} n_{i}=\frac{1}{\textit{w}(e^{(\epsilon_{i}-\mu)/kT})+\alpha}
    \eea
where, the function $\textit{w}(\zeta)$ satisfies the functional
equation
    \bea\label{w}
    \textit{w}(\zeta)^{\alpha}[1+\textit{w}(\zeta)]^{1-\alpha}=\zeta\equiv
    e^{(\epsilon-\mu)/kT}
    \eea
Equation (7) yields the correct solutions for the familiar bosons
$(\alpha=0)$, $\textit{w}(\zeta)=\zeta-1$ and fermions $(\alpha=1)$,
$\textit{w}(\zeta)=\zeta$. Some exact solutions of Equation (7) in
the special case of $\alpha$ have been presented in Aoyama
\cite{Aoyama}. But in the classical limit
$\exp[(\epsilon-\mu)/kT]\gg1]$,
    \bea
    \label{cl}\textit{w}(\zeta)=\zeta+\alpha-1,
    \eea
    \bea
    n_{i}=\frac{1}{e^{(\epsilon_{i}-\mu)/kT}+2\alpha-1}.
    \eea
we may evaluate the internal energy and particle number within
this limit,
    \bea
    U=\sum_{i}n_{i}\epsilon_{i}=\sum_{i}\frac{\epsilon_{i}}{e^{(\epsilon_{i}-\mu)/kT}+2\alpha-1}\\
    N=\sum_{i}n_{i}=\sum_{i}\frac{1}{e^{(\epsilon_{i}-\mu)/kT}+2\alpha-1}
    \eea
In the thermodynamic limit and two dimensional momentum space of
non-relativistic anyons with a mass $m$, the summation can be
replaced by the integral,
    \bea \sum_{i}\longrightarrow
    \frac{V}{h^{2}}2\pi m \int_{0}^{\infty} d\epsilon
    \eea
and, finally, the internal energy and particle number will be
    \bea
    U=\frac{2\pi V m}{h^{2}(1-2\alpha)}\int_{0}^{\infty}\frac{\epsilon
    d\epsilon}{[(1-2\alpha)e^{\mu/kT}]^{-1}e^{\beta\epsilon}-1},\\
    N=\frac{2\pi V m}{h^{2}(1-2\alpha)}\int_{0}^{\infty}\frac{
    d\epsilon}{[(1-2\alpha)e^{\mu/kT}]^{-1}e^{\beta\epsilon}-1},
    \eea
Or, in a more compact form,
    \bea\label{U1} U
    =\frac{V}{1-2\alpha}\Lambda^{-2}\beta^{-1}g_{2}(y),\\ \label{N1}N
    =\frac{V}{1-2\alpha}\Lambda^{-2}g_{1}(y),
    \eea
where, $\Lambda=\frac{h}{\sqrt{2\pi m kT}}$ is the mean thermal
wavelength of the particle, $\beta=1/kT$,
    \bea\label{B1}
    g_{l}(y)&=&\frac{1}{\Gamma(l)}\int_{0}^{\infty}\frac{x^{l-1}dx}{y^{-1}e^{x}-1}\nonumber\\
    &=&\sum_{n=1}^{\infty}\frac{y^{n}}{n^{l}}
    \eea
$\Gamma(l)$ here denotes the Gamma function, $y=(1-2\alpha)z$, and
fugacity is $z=e^{-\gamma}=e^{\mu/kT}$ \cite{Grad}.

%********************************************%
\section{Thermodynamic curvature of the anyon gas}
%********************************************%
Ruppeiner geometry is based on the entropy representation, where
we denote the extended set of $n+1$ extensive variables of the
system by $X=(U,N^{1},...,V,...,N^{r})$, while
 Weinhold worked in the energy representation in which the
extended set of $n+1$ extensive variables of the system were
denoted by $Y=(S,N^{1},...,V,...,N^{r})$. These variables are
identical to the extended set of extensive variables in the
entropy representation, except in the first slot where the
entropy, rather than the internal energy, appears. The
corresponding conjugate intensive parameters
    \bea P^{i}=\frac{\partial U}{\partial
    Y^{i}}
    \eea
are $P=(T,\mu^{1},...,-p,...,\mu^{r}).$ Then, the metrics of
Weinhold and Ruppeiner geometry are given by
    \bea
    d{s^{2}}_{R}=\partial_{i}\partial_{j}S \  dX^{i}dX^{j},
    \eea
and
    \bea
    d{s^{2}}_{W}=-\partial_{i}\partial_{j}U \ dY^{i}dY^{j}.
    \eea
In 1984, Mruga{\l}a\cite{Mrugala1} and Salamon et al. \cite{salamon}
proved that these two metrics are conformally equivalent with the
inverse of the temperature, $\beta$, as the conformal factor
    \bea
    d{s^{2}}_{R}=\beta \ d{s^{2}}_{W}.
    \eea
One can work in any thermodynamic potential representation that is
the legendre transform of the entropy or the internal energy. The
metric of this representation may be the second derivative of the
thermodynamic potential with respect
 to intensive variables; for example, the thermodynamic potential
$\Phi$ which is defined as,
    \bea
    \Phi=\Phi(\{F^{i}\})
    \eea
where, $F=(1/T,-\mu^{1}/T,...,P/T,...,-\mu^{r}/T)$. $\Phi$ is the
Legendre transform of entropy with respect to the extensive
parameter $X^{i}$,
    \bea
    F^{i}=\frac{\partial S}{\partial X^{i}}.
    \eea
The metric in this representation is given by
    \bea
    g_{ij}=\frac{\partial^{2}\Phi}{\partial F^{i}\partial F^{j}}.
    \eea
Janyszek and Mruga{\l}a used the partition function to introduce the
metric geometry of the parameter space
\cite{Mrugala2},
    \bea\label{M1} g_{ij}=\frac{\partial^{2}\ln
    Z}{\partial \beta^{i}\partial\beta^{j}}
    \eea where
$\beta^{i}=F^{i}/k$.

According to Equations (\ref{U1}) and (\ref{N1}), the parameter
space of an ideal anyon gas is $(1/kT,-\mu/kT)$ or equivalently
$(\beta,\gamma)$. For computing the thermodynamic metric, V
 is selected as the constant system scale. We can evaluate the elements of the  metric by the relevant definition in Equation ({\ref{M1}),
    \begin{gather}
    g_{\beta\beta}=\frac{\partial^{2}\ln Z}{\partial
    \beta^{2}}=-(\frac{\partial U}{\partial\beta})_{\gamma}=\frac{2B}{(1-2\alpha)}\beta^{-3}g_{2}(y),\nonumber\\
    g_{\beta\gamma}=g_{\gamma\beta}=\frac{\partial^{2}\ln Z}{\partial
    \beta\partial\gamma}=-(\frac{\partial U}{\partial
    \gamma})_{\beta}=\frac{2B}{(1-2\alpha)}\beta^{-2}g_{1}(y),\\
    \label{G1} g_{\gamma\gamma}=\frac{\partial^{2}\ln Z}{\partial
    \gamma^{2}}=-(\frac{\partial
    N}{\partial\gamma})_{\beta}=\frac{B}{(1-2\alpha)}\beta^{-1}g_{0}(y).\nonumber
    \end{gather}
where, $B=\frac{2m\pi V}{h^{2}}$ and
$y=(1-2\alpha)z=(1-2\alpha)e^{-\gamma}.$ From Equation (\ref{B1}),
one gets an important relation
    \bea
    \frac{\partial g_{l}(y)}{\partial
    y}=\frac{1}{y}g_{l-1}(y),
    \eea
It is easy to show that
    \bea
    \frac{\partial g_{l}(y)}{\partial
    \gamma}=-g_{l-1}(y).
    \eea

%%%(The Riemann and Ricci tensors  and the scalar curvature are
%%%respectively \begin{gather}\label{E1}
%%%R_{\alpha\beta\nu}^{\lambda}=\partial_{\beta}\Gamma_{\nu\alpha}^{\lambda}
%%%-\partial_{\nu}\Gamma_{\beta\alpha}^{\lambda}+\Gamma_{\nu\alpha}^{\eta}\Gamma_{\beta\eta}^{\lambda}
%%%-\Gamma_{\beta\alpha}^{\eta}\Gamma_{\nu\eta}^{\lambda}\nonumber\\
%%%Ric_{\mu\nu}= R_{\mu\lambda\nu}^{\lambda}\\
%%%R=g^{\mu\nu}Ric_{\mu\nu}\nonumber\end{gather} where
%%%$\Gamma_{\alpha\beta\eta}$ is the Christofeel symbol.

We consider systems with two thermodynamic degrees of freedom and,
 therefore, the dimension of the thermodynamic surface or parameter
space is equal to two ($D=2$). Thus, the scalar curvature is
given by
    \bea
    R=\frac{2}{\det g} R_{1212.}
    \eea

 Janyszek and Mruga{\l}a demonstrated
\cite{Mrugala3} that if the metric elements  are written purely as
the second derivatives of a certain thermodynamic potential, the
thermodynamic curvature may then be written in terms of the second
and the third derivatives. The sign convention for $R$ is
arbitrary, so $R$ may be either  positive or negative for any
case. Our selected sign convention is the same as that of Janyszek
and Mruga{\l} [9], but the different from \cite{Ruppeiner01}. In
two dimensional spaces, the formula for $R$ may be written as \bea
R=\frac{2\left|
         \begin{array}{ccc}
           g_{\beta\beta} & g_{\gamma\gamma} & g_{\beta\gamma} \\
           g_{\beta\beta,\beta} & g_{\gamma\gamma,\beta} & g_{\beta\gamma,\beta} \\
           g_{\beta\beta,\gamma} & g_{\gamma\gamma,\gamma} & g_{\beta\gamma,\gamma} \\
         \end{array}
       \right|}{{\left|
                  \begin{array}{cc}
                    g_{\beta\beta} & g_{\beta\gamma} \\
                    g_{\beta\gamma} & g_{\gamma\gamma} \\
                  \end{array}
                \right|}^{2}
       }.
    \eea
Using the following equations
    \begin{gather}
    g_{\beta\beta,\beta}=\frac{-6B}{(1-2\alpha)}\beta^{-4}g_{2}(y),\nonumber\\
    g_{\beta\beta,\gamma}=g_{\beta\gamma,\beta}=\frac{-2B}{(1-2\alpha)}\beta^{-3}g_{1}(y),\nonumber\\
    g_{\gamma\gamma,\beta}=g_{\beta\gamma,\gamma}=\frac{-B}{(1-2\alpha)}\beta^{-2}g_{0}(y),\nonumber\\
    g_{\gamma\gamma,\gamma}=\frac{-B}{(1-2\alpha)}\beta^{-1}g_{-1}(y).
    \end{gather}

we get
    \bea R=\frac{4\beta\Lambda^{2}}{V}(1-2\alpha)
    \{\frac{g_{0}(y){g_{1}}^{2}(y)-2g_{2}(y){g_{0}}^{2}(y)+g_{1}(y)g_{2}(y)g_{-1}(y)}
    {[2g_{2}(y)g_{0}(y)-{g_{1}}^{2}(y)]^{2}}\}.
    \eea

 In Table \ref{table1}, we have collected some numerical values
of $R$ computed by Maple. $R$ is given in units of
$\frac{4\beta\Lambda^{2}}{V}$ and $1/kT=constant$, i.e. for an
isotherm.
    \begin{table}
    \begin{center}
    \begin{tabular}{|c||cccc|}
    \hline
    $ $ &  $z=0.001$ & $z=0.005$ & $z=0.01$ & $z=0.1$ \\
    \hline\hline
    $\alpha=0.0$&  0.2500144897  & 0.2500699315  & 0.2501403817  & 0.2515411180 \\
    $\alpha=0.1$&  0.2000094605  & 0.2000447381  & 0.2000895695  & 0.2009653621 \\
    $\alpha=0.2$&  0.1500051368  & 0.1500250813  & 0.1500503648  & 0.1505316710  \\
    $\alpha=0.3$&  0.1000016230  & 0.1000111807  & 0.1000223691  & 0.1002314415  \\
    $\alpha=0.4$&  0.0500012375 & 0.0500028979  & 0.0500055903  & 0.0500566821  \\
    $\alpha=0.5$&  0.0           & 0.0           & 0.0           & 0.0           \\
    $\alpha=0.6$&  -0.0499987375 & -0.0499971990 & -0.0499944358 & -0.0499455377 \\
    $\alpha=0.7$&  -0.0999977160 & -0.0999888716 & -0.0999778088 & -0.0997863640 \\
    $\alpha=0.8$&  -0.1499951380 & -0.1499750132 & -0.1499502686 & -0.1495285050 \\
    $\alpha=0.9$&  -0.1999909479 & -0.1999556176 & -0.1999117788 & -0.1991775621 \\
    $\alpha=1.0$&  -0.2499859954 & -0.2499310365 & -0.2498624835 & -0.2487388188 \\
    \hline
    \end{tabular}
    \end{center}
    \bigskip
    \caption{Scalar thermodynamic curvature $R$ for chosen value of
    the fugacity (in classical limit) and various value of
    $0\leq\alpha\leq1$ for anyon gas. }\label{table1}
    \end{table}
It is evident from Table \ref{table1} that for
$\alpha<\frac{1}{2}$, the thermodynamic curvature $R$ is always
positive while it is always negative for $\alpha>\frac{1}{2}$.
This result indicates that the anyon gas is more stable when
$\alpha>\frac{1}{2}$.
 For $\alpha=\frac{1}{2}$, the
thermodynamic curvature is zero. So, the sign of $R$ changes at
$\alpha=\frac{1}{2}$. In the Calssical limit, it has been shown that
\cite{Wu},
    \bea
    PV=NkT[1-(1-2\alpha)N\Lambda^{2}/4V],
    \eea
So the "statistical interactions" are attractive or repulsive
depending on whether $\alpha<\frac{1}{2}$ or $\alpha>\frac{1}{2}.$
Therefore, the thermodynamic curvature is positive for attractive
statistical interactions and it is negative for repulsive
statistical interactions. Our interpretation of stability is,
therefore, consistent with bosonic and fermionic gases. This
interpretation measures the looseness of the system to
fluctuations and does not refer to the fact that the metric is
definitely positive. For $\alpha=\frac{1}{2}$, the equation of
state is like that of an ideal classical gas where its
thermodynamic curvature is zero.

%%%%%%%%%%%%%%%%%%%%%%%%%%%%%%%%%%%%%%%%%%%%%%%%
\section{Beyond the classical limit}
%%%%%%%%%%%%%%%%%%%%%%%%%%%%%%%%%%%%%%%%%%%%%%%%
In the last section, the thermodynamic curvature was evaluated in
the classical limit. In what follows, we will first investigate a
small deviation from the  classical limit and its results for the
thermodynamic curvature. Deviations from the classical limit and
a more general solution of Eq. (8) is given by the following
function:
    \bea
    \textit{w}(\zeta)=\zeta+\alpha-1+\frac{c_1}{\zeta}+\frac{c_2}{\zeta^{2}}
    +\frac{c_3}{\zeta^3}+\cdots,
    \eea
where the constant coefficients $c_1$, $c_2$, $\cdots$ can be
evaluated on the condition that at each order of $\zeta $,  the
$\textit{w}(\zeta)$ satisfies (\ref{w}) and so we get,
    \bea
    c_1&=&\frac{1}{2}\alpha(1-\alpha),\nonumber\\
    c_2&=&\frac{1}{3}\alpha(1-\alpha)(1-2\alpha),\\
    c_3&=&\frac{1}{8}\alpha(1-\alpha)(1-3\alpha)(2-3\alpha),\nonumber
    \eea
The other coefficient can be evaluated along the same lines. For a
small deviation from the classical limit,  we may  use only the
first correction in (35) which leads to the following form for
Eq. (7):
    \bea
    n_{i}=\frac{1}{\zeta+2\alpha-1+\frac{c_1}{\zeta}}.
    \eea
By expanding about the classical value of $n_{i}$, we can find a
correction up to the leading order:
    \bea
    n_{i}=\frac{1}{\zeta+2\alpha-1}-\frac{c_1}{\zeta^3}.
    \eea
Now, we can easily evaluate the internal energy and particle
number as well as the thermodynamic metric along the lines set in
 the previous section so that  finally, the
thermodynamic curvature could be worked out numerically. The
results are represented in Fig. (\ref{f1}).
    \begin{figure}
     \center\includegraphics[width=8cm]{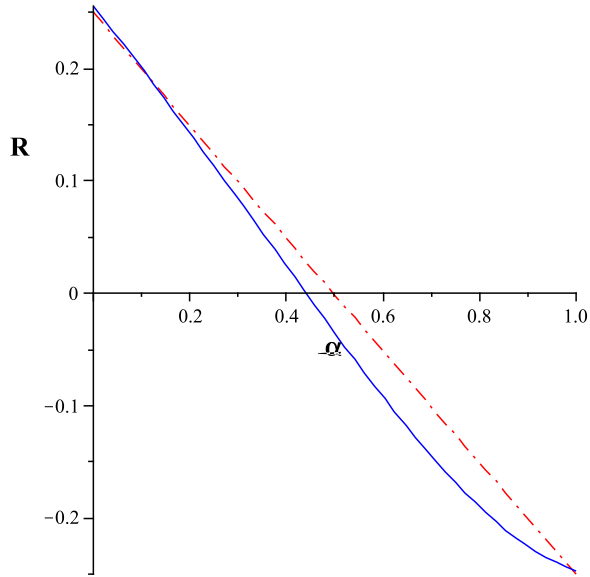}\\
     \caption{The thermodynamic curvature as a function of $\alpha$.
     The dash-dot line corresponds to the  thermodynamic curvature
     for the
     classical limit for an isotherm and $z=0.01$.
      The solid line corresponds to the small deviation from the
      classical limit and represents the thermodynamic curvature
      at $z=0.3$. }\label{f1}
    \end{figure}
It can be seen that the values of the thermodynamic curvature are
different from the classical limit. It is interesting that the
zero point of the thermodynamics curvature is shifted from
$\alpha=\frac{1}{2}$ (classical limit) to the lower numbers. This
means that quantum corrections change the value of $\alpha $
where we have a free non-interacting gas. This is the basic
result of this paper.

The thermodynamic curvature for $\alpha=\frac{1}{2}$ can be worked
out in the full physical range. For this special case,  Eq. (8)
becomes a quadratic equation which can be easily solved to give:
    \bea
    n_{i}=\frac{1}{\sqrt{\frac{1}{4}+\zeta^2}}.
    \eea
Thus, the internal energy and particle number can be obtained,
    \bea
    U&=&\frac{8\pi
    V m z\beta^2}{\hbar^2}\,{_{3}F_{2}}\left(\frac{1}{2},\frac{1}{2},\frac{1}{2};\frac{3}{2},\frac{3}{2}
    ;-\frac{z^2}{4}\right)\nonumber\\
    N&=&\frac{4\pi
    Vm\beta}{\hbar^2}{\it arcsinh} \left(\frac{z}{2}\right).
    \eea
Calculation of the thermodynamic curvature is straightforward and
 the result is represented in Fig. (\ref{f2}). It shows that for
$\alpha=\frac{1}{2}$, the thermodynamic curvature is zero only at
the classical limit.
\begin{figure}
  \center\includegraphics[width=8cm]{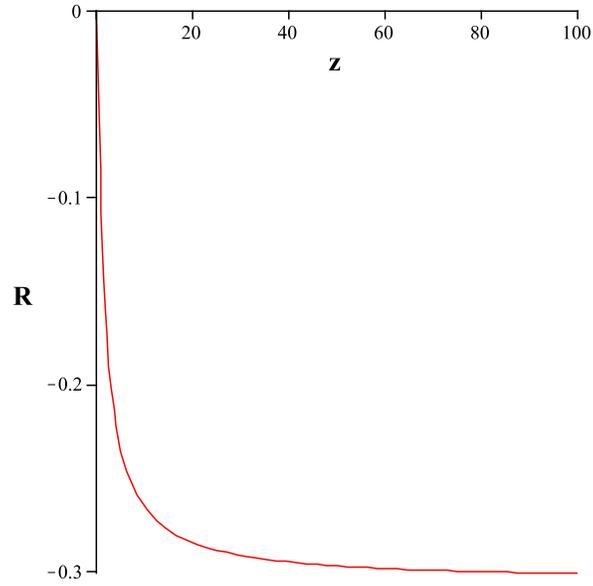}\\
  \caption{The thermodynamic curvature as a function of z (fugacity) for
  $\alpha=\frac{1}{2}$  and an isotherm in the full physical range.}\label{f2}
\end{figure}

%%%%%%%%%%%%%%%%%%%%%%%%%%%%%%%%%%%%%%%%%%%%%%%%
\section{Conclusion}
%%%%%%%%%%%%%%%%%%%%%%%%%%%%%%%%%%%%%%%%%%%%%%%%
The ideal anyonic gas in the Calssical limit has two different
behaviour depending on whether $\alpha<\frac{1}{2}$ or
$\alpha>\frac{1}{2}.$ For $\alpha<\frac{1}{2}$, the statistical
interaction is attractive and the scalar thermodynamic curvature
is positive. Thus, we may call the ideal anyonic gas in this case
"Bose-like". Along these lines,  the statistical interaction is
repulsive for $\alpha<\frac{1}{2}$ and the scalar thermodynamic
curvature is negative. Thus, the ideal anyonic gas in this case is
"Fermi-like". As shown in \cite{Mrugala2,Janyszek}, we may
consider the thermodynamic curvature as a measure of the
stability of the system: the bigger the value of $R$, the less
stable is the system. Therefore, the ideal anyonic gas in a
Fermi-like case ($\alpha>\frac{1}{2}$) is more stable than in the
Bose-like case ($\alpha<\frac{1}{2}$). Deviations from the
classical limit move the zero point of the thermodynamic
curvature from $\alpha=\frac{1}{2}$ to the lower values.

%%%%%%%%%%%%%%%%%%%%%%%%%%%%%%%%%%%%%%%%%%%%%%%%%%%%%%%%%%%%%%%%%%%%%%%%%%%%%%%%%%

\end{document}